\documentclass{PoS}
\usepackage[utf8]{inputenc}
\usepackage{caption}
\newcommand{\be}{\begin{equation}}
\newcommand{\ee}{\end{equation}}
\newcommand{\bea}{\begin{eqnarray}}
\newcommand{\eea}{\end{eqnarray}}

\title{The meson spectrum of large N gauge theories}

\ShortTitle{ The meson spectrum of large N gauge theories}

\author{Margarita García Pérez \\ 
       Instituto de Física Teórica UAM/CSIC, Nicolás Cabrera 13-15,\\
       E-28049 Universidad Autónomade Madrid, Madrid, Spain\\
        E-mail: \email{margarita.garcia@uam.es}}

\author{\speaker{Antonio González-Arroyo}\\
        Departamento de Física Teórica, Módulo 15, Cantoblanco,\\
	E-28049 Universidad Autónoma deMadrid, Madrid, Spain\\
	and\\
	Instituto de Física Teórica UAM-CSIC, Nicolás Cabrera 13-15,\\
	E-28049 Universidad Autónomade Madrid, Madrid, Spain\\
        E-mail: \email{antonio.gonzalez-arroyo@uam.es}}

\author{Masanori Okawa  \\
      Graduate School of Science, Hiroshima University,\\
      Higashi-Hiroshima 739-8526, Japan \\
      E-mail: \email{okawa@hiroshima-u.ac.jp}}

\abstract{We present our preliminary results on the  determination of the low lying meson spectrum
for pure gauge theory in the large $N$ limit. Some results are also
shown for the theory with two flavours of quarks in the adjoint
representation. }

\FullConference{37th International Symposium on Lattice Field Theory - Lattice2019\\
		16-22 June 2019\\
		Wuhan, China}

\begin{document}

\section{Introduction}
The study of gauge theories in the large $N$ limit is of fundamental
importance for the understanding of the dynamics of these theories,
which are at the core of the Standard Model and many of its extensions. 
The advantage of the large $N$ limit rests in its simplicity, which
comes at no price regarding the richness of phenomena encompassed by
these theories. The simplification appears already in perturbation
theory, since only planar diagrams survive the limit. Large $N$ gauge 
theories are also sitting at the crux of several new methodologies 
emerging from string theory, such as
holography. Some results obtained by these new methods refer to
non-perturbative quantities as the meson spectrum and the string
tension. This provides a challenge to lattice gauge theories (LGT), which is
the most solid first principles approach to the study of the
non-perturbative behaviour of gauge theories. Hence, we believe that,
besides the obvious interest of computing non-perturbative observables
of the standard model with direct phenomenological impact,  a
thorough study of large $N$ gauge theories should be addressed. This
should be done within the standards of present day LGT, where all
errors, statistical and systematic, are controllable and estimated.
This is the program that we have set ourselves to accomplish. We have
already obtained some results in this respect~\cite{string_tension} and the
present talk reports on the present status of the calculation of the
lowest lying meson spectrum. A complete set of results on masses and
decay constants will appear soon~\cite{prep}.

\section{Methodology}
The most important tool used in obtaining non-perturbative results on
the lattice is the Monte Carlo method. As such, it demands dealing with
a large but finite number of degrees of freedom. This implies a finite
lattice volume, which indirectly also provides  a minimal value of the
lattice spacing at which the results are not seriously affected by the
finite physical volume. Fortunately, these questions are part of the
daily procedures of the LGT community and we know how to test and
estimate the errors involved, by performing measurements at various
values of the lattice volume and the lattice spacing.  For the
case of large $N$ gauge theories the finite volume does not suffice and 
one needs to restrict to finite values of $N$ and extrapolate the
results. This is the standard methodological procedure that many
authors have used to obtain non-perturbative values in the large $N$
limit. From that perspective large $N$ does not provide a
simplification of lattice gauge theories. 

Fortunately, there is indeed a simplification that takes place in the
large $N$ limit within the lattice approach. This follows from the
observation of Eguchi and Kawai~\cite{EK} who argued that finite
volume corrections go to zero in the large $N$ limit. Indeed, in its
more standard  version, the proposal does not hold. There are several
alternatives proposed over the years to transform this idea into a
reality (We refer the reader to our previous and forthcoming
publications for a full list of references). The  method that we
are using is based on  a fairly slight modification of the original proposal of Eguchi
and Kawai, which was introduced by two of the present
authors~\cite{TEK1,TEK2,TEK3}. The idea is to use `t Hooft twisted
boundary  conditions (TBC) instead of purely periodic ones. TBC can be easily incorporated to the lattice
regularization~\cite{GJK}. Furthermore, as observed in ~\cite{TEK1},  
Eguchi-Kawai  arguments are valid for  TBC as well, but the
main assumptions involved in the vanishing of finite volume
corrections should work better for well-chosen twist fluxes.

The basis of our methodology is the statement that physical observables 
in the large $N$ limit, on the lattice and in the continuum, take the same value 
irrespective of the lattice volume, provided the limit is taken with
appropiately chosen twisted boundary conditions. Although, not at all
compulsory, we will take the extreme case of reducing the volume to  
a 1-point lattice, defining a matrix model known
as the Twisted Eguchi-Kawai model (TEK)~\cite{TEK2}.   The partition function
of this model,
which is used  to generate gauge field configurations, is given by:
\be
\label{TEK}
Z_{\mathrm{TEK}}= \prod_\mu \left( \int d V_\mu \right)  \exp\{b N 
\sum_{\mu \ne \nu} z_{\mu  \nu} \mathrm{Tr}( V_\mu V_\nu V_\mu^\dagger
V_\nu^\dagger )\} 
\ee
The $V_\mu$ are 4 SU(N) matrices integrated with the Haar measure.
The constant $b$ is the lattice version of the inverse `t Hooft
coupling, that is kept fixed when taking the large $N$ limit. 
The constants $z_{\mu  \nu}\in \mathrm{Z(N)}$ are precisely the fluxes
that appear in the definition of the twisted boundary conditions
($z_{\mu  \nu} = z_{\nu  \mu}^*$). To
ensure the validity of the reduction mechanism one should make an
appropriate choice. The original Eguchi-Kawai model was based in
taking $z_{\mu  \nu}=1$, which has problems at weak coupling where the
continuum limit is taken~\cite{QEK}. There are many possible choices but our
prefered choice is the one that treats all lattice directions in the  
most symmetric way. For that purpose one should take $N$ to be the
square of an integer $N=\hat{L}^2$ and $
z_{\mu  \nu}= \exp\{2 \pi i k_N \epsilon_{\mu \nu} /\hat{L}\}$, 
where $\epsilon_{\mu \nu}=1$ for $\mu<\nu$ and $k_N$ is an integer
defined modulo $\hat{L}$ and coprime with it. 

The main formula  underlying the reduction mechanism is given
by
\be
\label{reduction}
\lim_{N \longrightarrow \infty}  \lim_{V \longrightarrow \infty}
\frac{1}{N} \langle \mathrm{Tr}(U(\gamma)) \rangle =  \lim_{N \longrightarrow
\infty} \frac{z(\gamma)}{N}  \langle \mathrm{Tr}(V(\gamma)) \rangle_{\mathrm{TEK}} \equiv
W(\gamma)
\ee
where the left-hand side is the  standard Wilson loop expectation
value at infinite volume and infinite $N$: $W(\gamma)$. The second expression
is the expectation value of the equivalent loop in the TEK model,
where $V(\gamma)$ is the corresponding ordered product of the position
independent $V_\mu$ matrices and $z(\gamma)$ a path dependent element
of Z(N) (to be specified later). 
The equality holds at every value of the
coupling $b$, if the same action (Wilson action in our case) is taken
for the infinite volume and the reduced model.

The validity of Eq.~\ref{reduction}  for our flux choices has been 
verified in various ways. First by checking the validity of
the assumptions entering   the non-perturbative proof of Eguchi and Kawai.
The result can also be  proven analytically 
to all orders of perturbation theory~\cite{TEK2}. Finally,  by direct lattice
computations~\cite{testing} of both sides of the
equation  for  various lattice sizes and values of $N$.

For theoretical and practical purposes it is important to understand
the nature of the finite $N$ corrections to
Eq.~\ref{reduction}. The perturbative proof~\cite{TEK2} gives us information about this.
At large $b$ the reduced model is driven to a mimimum of the TEK
action. This is given by  matrices $V_\mu=\Gamma_\mu$, where the new
matrices, called twist-eaters, satisfy
$\Gamma_\mu \Gamma_\nu= z_{\nu \mu} \Gamma_\nu \Gamma_\mu$. 
Indeed, for a given closed lattice path $\gamma$, the product of the
twist-eaters along the path $\Gamma(\gamma)$ is just the unit matrix
times the $z^*(\gamma)$ factor mentioned earlier.  Perturbing around 
these minima we generate a perturbative expansion where the planar
 diagrams coincide with those obtained for a lattice volume of size  
 $\hat{L}^4$. This explains the volume independencein the large $N$ limit,
 and shows that these finite $N$ corrections adopt the form of finite
volume corrections.   Non-planar diagram suppression is different
than in ordinary infinite volume large $N$ theory and depends on the
choice of $k_N$. Indeed, the choice of $k_N$ has a strong effect on
the validity of the reduction program and here we will stick to the
prescription given in Ref.~\cite{TEK3}. 


Adding a few flavours of quarks in the fundamental representation can
be readily done in the standard lattice gauge theory. In the large $N$
limit, quarks do not affect the gauge field probability distribution.
Quark and meson propagators are computed in the background field of
the pure gauge configurations. The reduced model is not constructed on
the basis of a discretized lattice gauge theory with  quarks  reduced
to a single point. This would lead to problems, as TBC become singular
for quarks in center-sensitive representations. The philosophy is
quite different. We can in principle allow the quarks to live in an
infinite lattice,  but they propagate in the background field obtained
from the $V_\mu$. The situation is similar to that encountered in
condensed matter theory in which electrons propagate in an infinite
solid, but the background field produced by the ions has the
periodicity of the lattice. The infinite lattice gives rise to a
continuous Bloch momentum, and the energy levels depend on them giving
rise to bands. The actual construction is long enough to
make it impossible to review here.  We refer the reader to our previous
papers in which the method is explained~\cite{mesons}. We should just
mention that the methodology can be implemented for different types of
lattice fermions. In this work we have used Wilson fermions as before,
but also included results with twisted mass. 

To conclude this section let us comment briefly about systematic
errors. Equation~\ref{reduction} holds at infinite $N$, while our
simulations have been done at  rather large, but finite,  values of $N$.
A good deal of this 
finiteness is equivalent to working on a finite volume of size
$\hat{L}^4$. Thus, the corresponding physical period  of the torus is
given by $l=\hat{L}a(b)$, where $a(b)$ is the lattice spacing in
physical units. The values that we use in this work are given in units
of the  string tension, as determined from our previous
work~\cite{string_tension}. 
If we want our results not to be affected, this finite effective size
should be kept much larger than the relevant physical scales of the
problem. No doubt that one of these scales is $\Lambda_{QCD}$. This
forces $\hat{L}=\sqrt{N}$ to be large enough and limits the maximum
value of $b$ as well. This is the standard restriction of LGT, with
$\sqrt{N}$ replacing the lattice  length $L$. When approaching the
chiral limit the pion mass vanishes and the pion propagator gets
affected by the finite volume. Hence, we have also avoided coming too
close to this limit. All our results have been for values of $b$ and
$N$ for which the effective lattice size is kept within reasonably
safe limits. 

A final comment about the comparison of our method and that based on
extrapolation from a more standard lattice approach, as used in the
Refs.~\cite{Bali,degrand}.   In some sense the
methods can be seen as  complementary. Our method goes directly to the
large $N$ limit, but does not produce the correct finite $N$
corrections, so a combination of both results could stabilize the
extrapolation and provide better determined $1/N^2$ corrections. Some
of the problems appearing at small  $N$ such as chiral logs, are
absent in our method. Ultimately, all methodologies should agree on
the results. In any case, we think that all works performed at finite
lattices should consider the use of TBC, which would reduce
considerably the finite size errors at almost no cost.

\section{Results}

This work reports the calculation of the lowest lying meson masses
using our procedure based on the TEK model. The gauge field is
generated with the TEK probability distribution using the
overrelaxation technique explained in Ref.~\cite{overrelaxation}. The
results presented here were obtained with $N=289$, implying
$\hat{L}=17$. In order to check and study the continuum limit we
simulated three values of $b$, 0.355, 0.36 and 0.365. For these values,
the lattice spacing measured in string tension units is given by
0.241, 0.206 and 0.178 respectively, with errors of order 2 \%. This
corresponds to  effective lattice sizes of  4.10, 3.50 and 3.03. 
Our results are based on 800
configurations per value of $b$. We have
also generated results for b=0.37, but  some observables might
start to exhibit finite $N$ effects. Our complete  results
including b=0.37 and various values of $N$ will be presented in our
future publication~\cite{prep}. 

The methodology to extract masses is based on measuring exponential
decay of the correlation functions of operators with the right quantum
numbers. The basic operators are those obtained by inserting a
Clifford algebra matrix $\Gamma$ inside a local $\bar\Psi(x) \Gamma
\Psi(x)$ meson operator. This gives the $J^{PC}$ quantum numbers
$0^{-+}$, $1^{--}$,
$0^{++}$  $1^{++}$ and $1^{+-}$. We will refer to these states with the names given in
standard QCD ($\pi$, $\rho$, $a_0$, $a_1$ and $b_1$).
For each possible matrix we consider a whole family of operators (up
to 12)  obtained by smearing the gauge field and the fermion field. We then
use a variational technique to generate an optimal operator which
couples maximally to the lowest mass state with the corresponding
quantum numbers. 

As mentioned earlier, most of our results are obtained for Wilson
fermions. We used between 5 and 7 different values of the hopping
parameter for each $b$. For the pseudoscalar and vector cases, which
we label $\pi$ and $\rho$, we also studied the twisted mass Dirac
operator for 4 values of $\mu$. The use of twisted mass is specially
important for the determination of the decay constants and
operator renormalization $Z$ coefficients, which will not be presented
here. In summary  the main results of our investigation are the
following:\\
{\bf 1.} Good exponential fall-off of the meson correlators is observed,
allowing a fairly precise determination of the masses. Examples are
shown in Fig.~1.\\
{\bf 2.} The pseudoscalar propagator behaves as expected for a
spontaneously chiral symmetry breaking phase. This means that the mass
of the lightest state squared vanishes linearly when the hopping $\kappa$
parameter tends to a certain value $\kappa_c$. For the twisted mass
case the vanishing is linear with the parameter $\mu$ as shown in
Fig.~2. \\
{\bf 3.} The masses of the other states depend linearly on $m_{\mathrm
PCAC}$. In Fig.~3 we display the results for
the $\rho$ and $a_0$ states. In all cases the masses in string tension
units are consistent  for the two largest values of $b$.  A
simultaneous linear fit to these two values gives the slopes and
intercepts given in Table 1. Both statistical and systematic errors
are given. The intercepts correspond to the masses
of these mesons in the chiral limit. Our results are compared with the
values of Ref.~\cite{Bali}. \\ 
{\bf 4.} Our data shows good scaling behaviour within our statistical
errors.  This can be easily appreciated from Fig.~4 which
displays the masses in the chiral limit obtained by fitting
independently the results for each value of $b$. The horizontal bands
give our estimates for the masses in the continuum limit. \\
{\bf 5.} We have performed a full analysis of systematic errors. We
address the reader to our future publication giving full details of the
methodology, tests and results.  

\begin{figure}
\begin{tabular}{cc}
\hspace*{-1cm}  \includegraphics[width=0.5\textwidth]{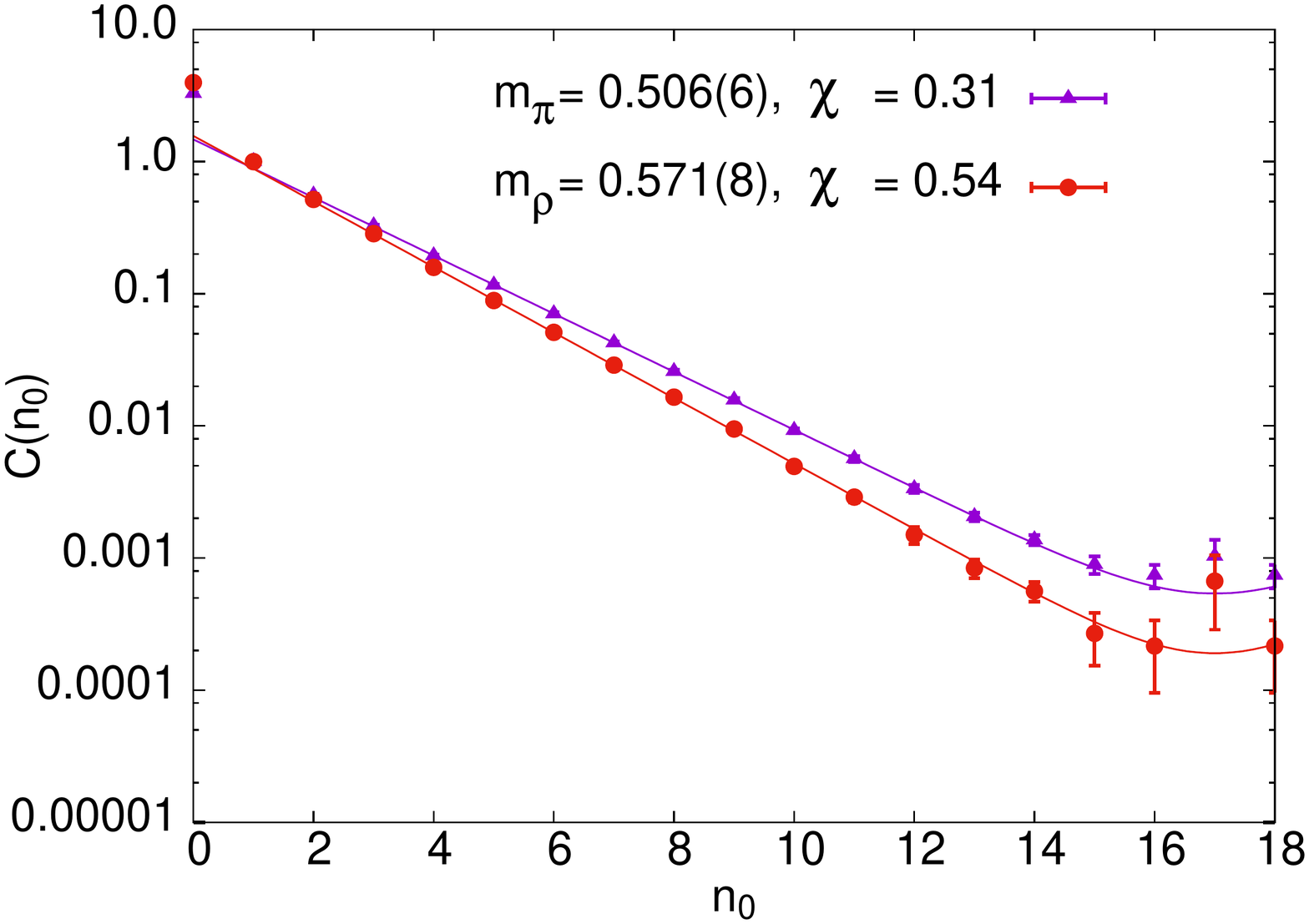}
  \label{correlator}
    &
      \includegraphics[width=0.45\textwidth]{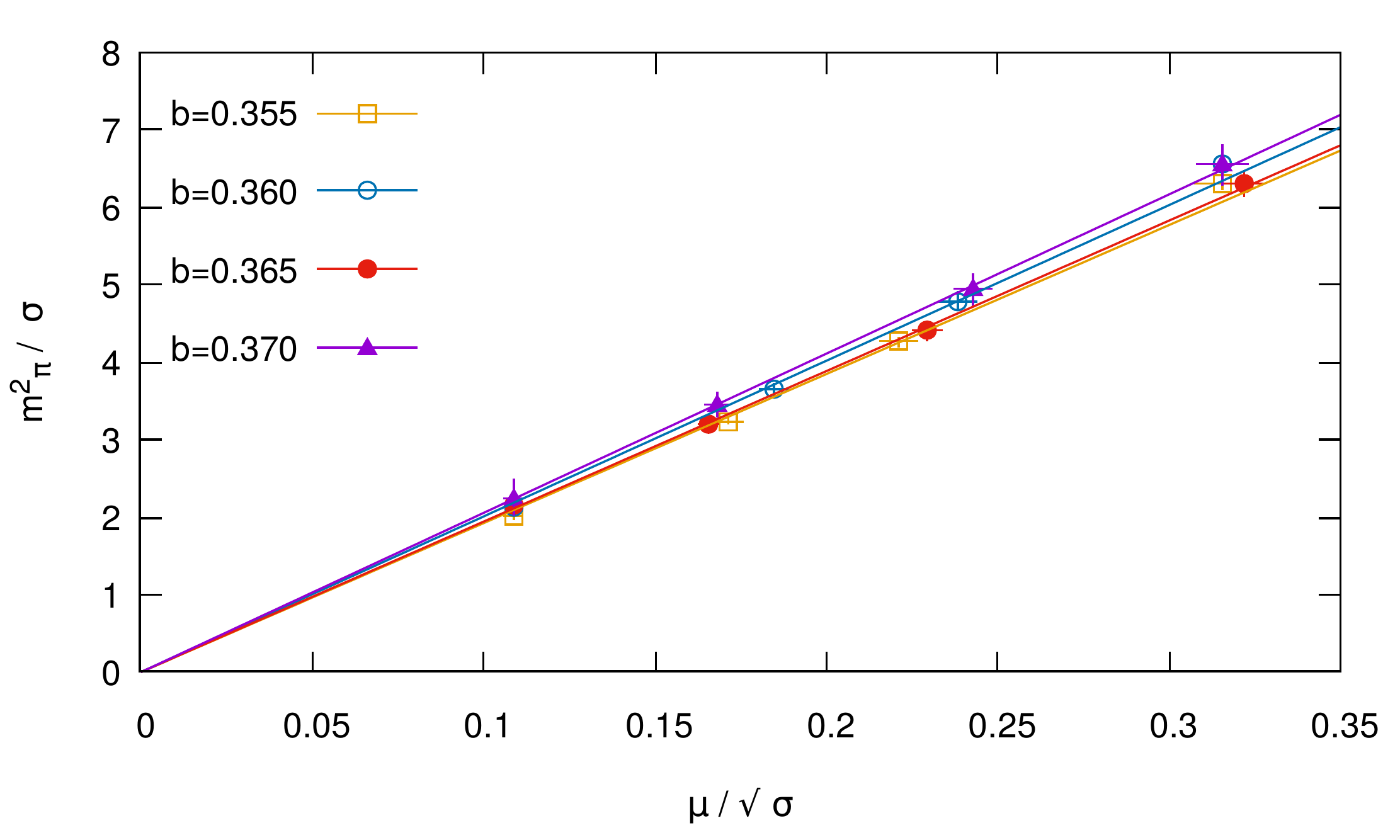}\label{Fig2}
         \\
            Fig. 1: Example of pion and $\rho$ correlators  &  Fig. 2: $m_\pi^2$  versus  $\mu$ for
	    twisted mass fermions.\\ for the  optimal operator. & 
             \end{tabular}
\end{figure}

\begin{figure}
\begin{tabular}{cc}
\hspace*{-1cm}  \includegraphics[width=0.45\textwidth]{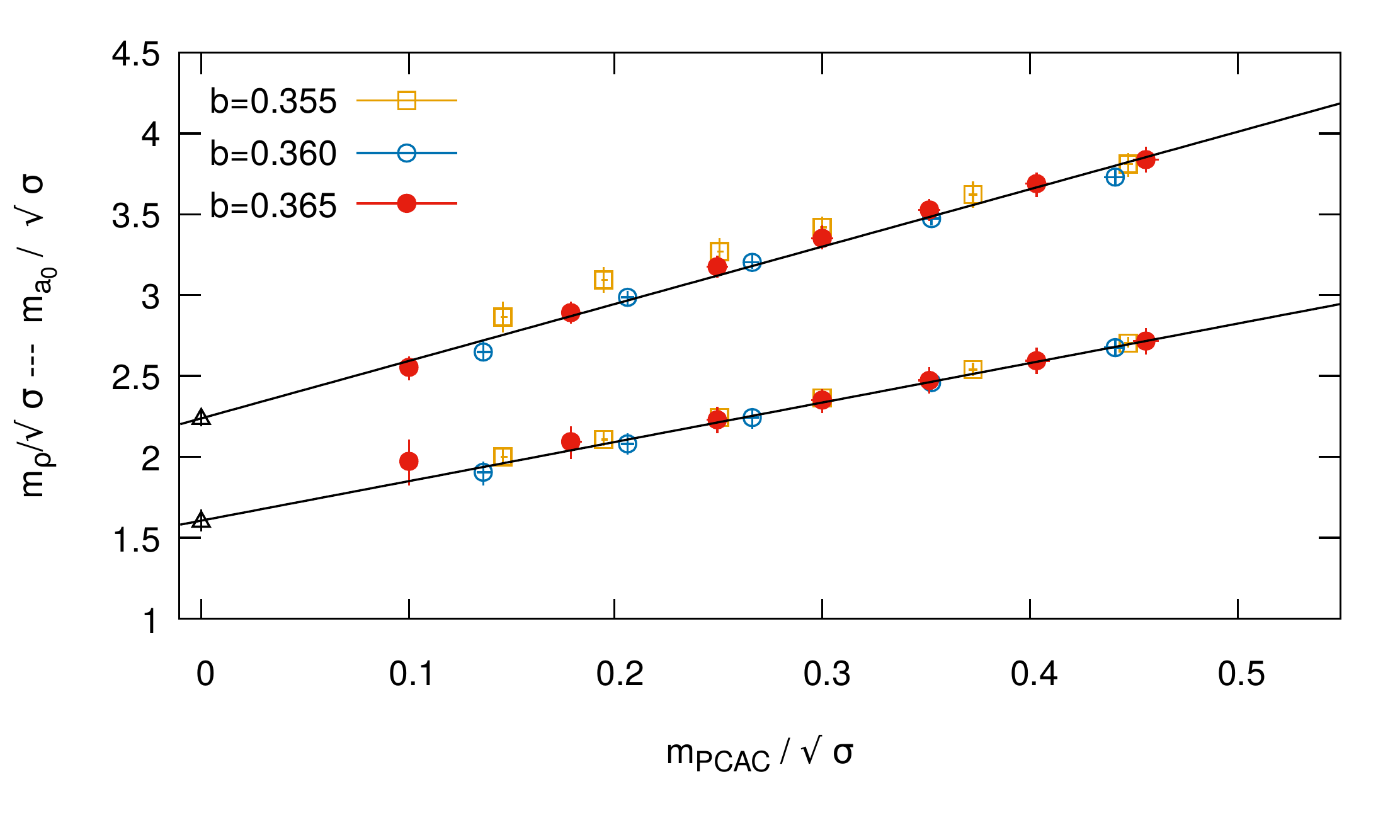}
  \label{Fig3}
    &
      \includegraphics[width=0.45\textwidth]{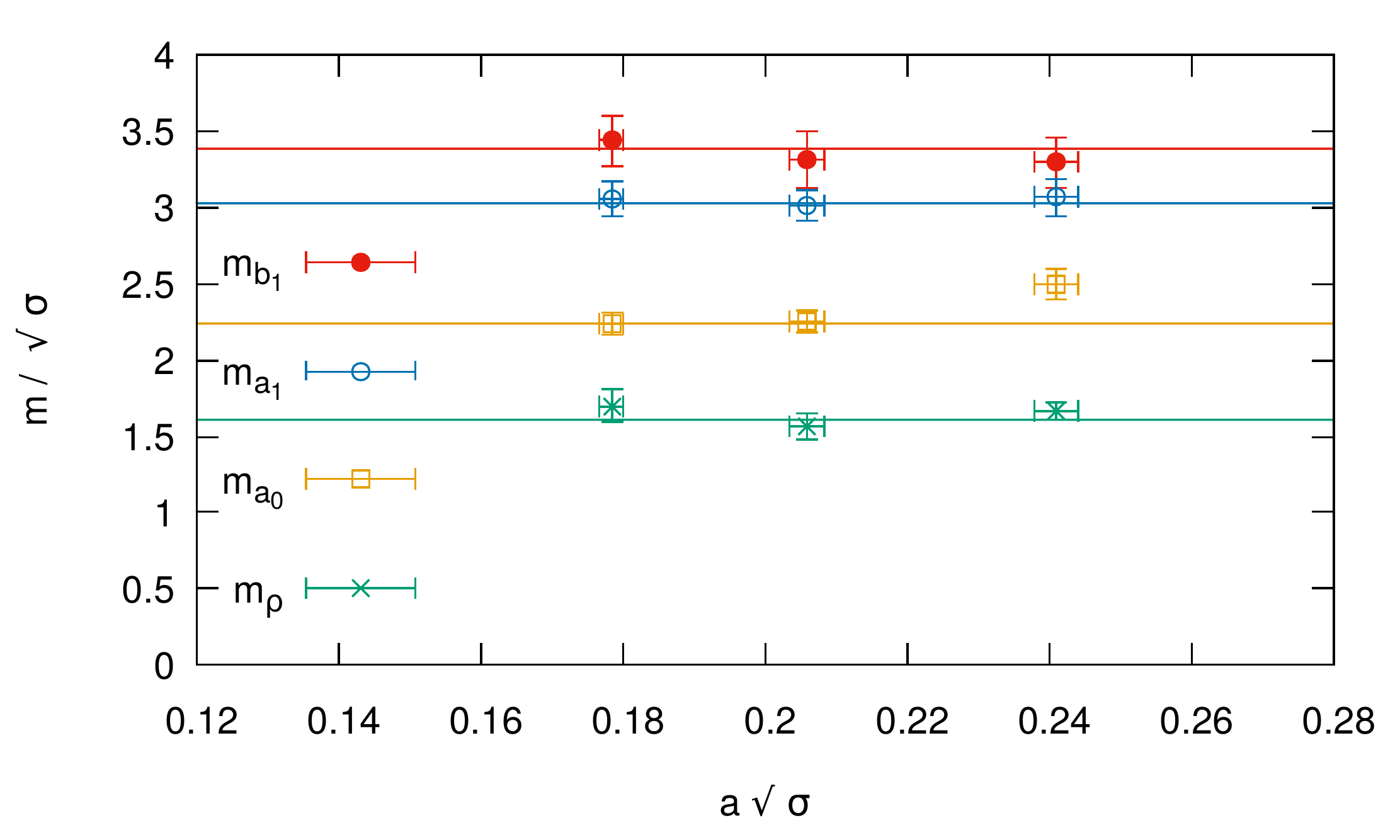}\label{Fig4}
         \\
            Fig. 3: Mass of $\rho$ (lower) and $a_0$ mesons (upper)
	    &  Fig. 4:
	    Masses in the chiral limit. \\ in units of  the
	                string tension versus $m_{\mathrm{PCAC}}$. & \\
             \end{tabular}
\end{figure}

\begin{figure*}
\begin{minipage}{0.35\linewidth}
\label{tab1}
\begin{tabular}{| c |l | l | l|} \hline
\hspace*{0.5 cm}& slope& mass/$\sqrt{\sigma}$ & Ref.~\cite{Bali}\ \\ \hline
$\rho$ & 2.43(12)& 1.61(7)(5) & 1.538(7)\\ \hline
$a_0$ & 3.53(22) & 2.24(5)(4) & 2.40(4)\\ \hline
$a_1$ & 2.32(15) & 2.99(8)(2) & 2.86(2)\\ \hline
$b_1$ & 2.22(17)& 3.20(12)(18) & 2.90(2)\\ \hline
  \end{tabular}
\captionof{table}{\hspace*{-3.3 mm} Summary of meson masses.}
  \end{minipage}
\hfill 
\begin{minipage}{0.5\linewidth}
  \includegraphics[width=\textwidth]{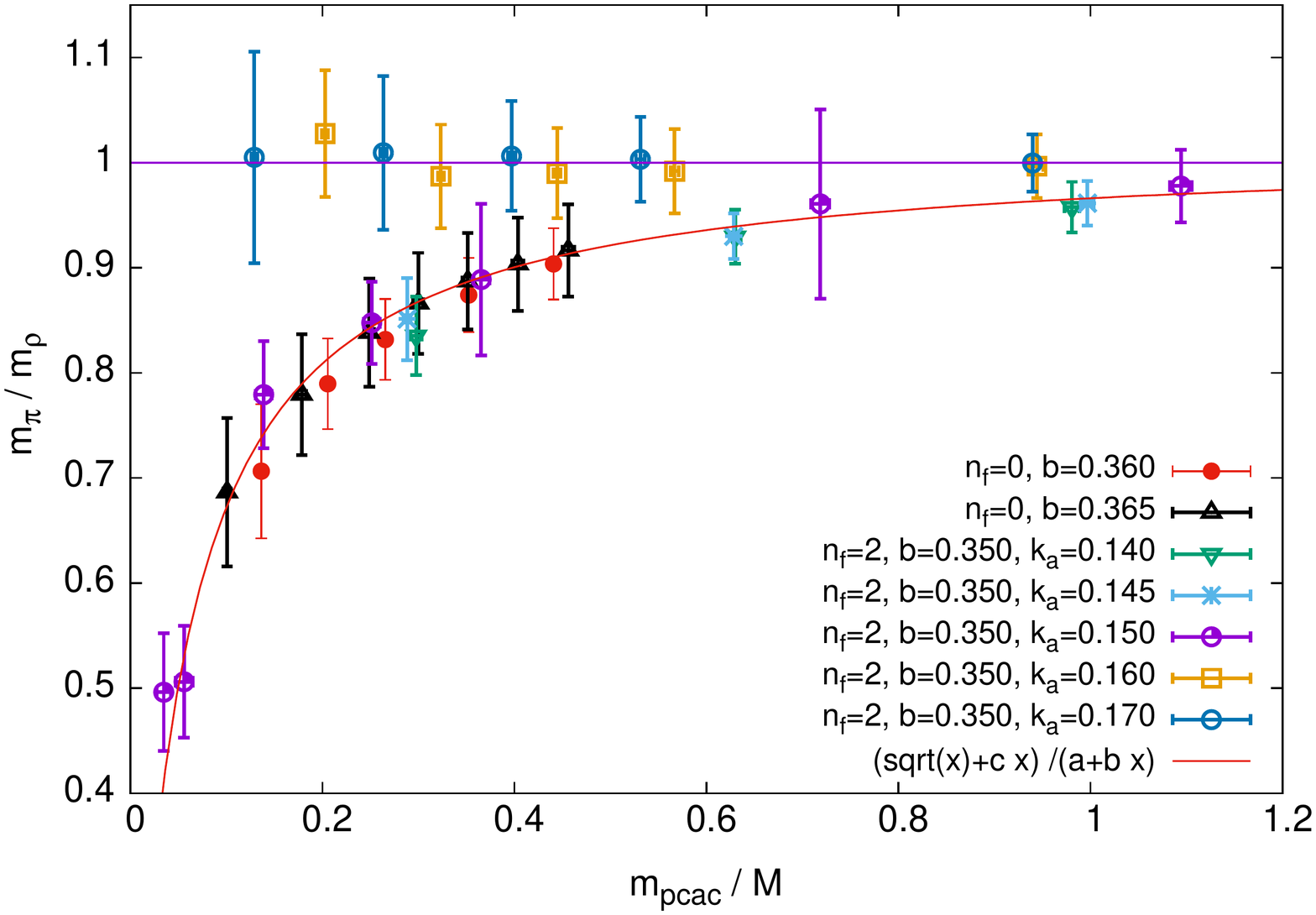}
\caption*{Fig. 5: The ratio of masses of $\pi$ and $\rho$ mesons \\
made of fundamental quarks of mass $m_{\mathrm{PCAC}}$ for \\
2 flavours of
adjoint quarks with various quark masses. }
\label{Fig5}
\end{minipage}
 \end{figure*}



\section{Conclusions and Outlook}
Our results show that the idea of volume reduction can be used
succesfully to determine the physical properties of large $N$ gauge
theories. We want to stress that this can be extended to theories with
dynamical fermions. This includes theories with quarks in the adjoint
representation, encompassing very interesting theories ranging from
SUSY Yang-Mills to theories that are expected to lay within the
conformal window. In Fig.~5 we show results for the ratio of pion to
rho masses (made of quarks in the fundamental) for the theory 
with $N_f=2$ light quarks in the adjoint, in deep contrast with the
chiral symmetry breaking of the $N_f=0$ theory.  Furthermore, theories with
quarks in the fundamental representation in the Veneziano limit also
seem to be accesible with our methods.

\acknowledgments 

A.G-A and M. G.P. acknowledge financial support from the MINECO/FEDER grant
FPA2015-68541-P, the MINECO Centro de Excelencia Severo Ochoa Program
SEV-2016- 0597 and the EU H2020-MSCA-ITN-2018-813942 (EuroPLEx).
 M.O. is supported by the Japanese MEXT grant No. 17K05417.
 This research used computational resources of the SX-ACE system
 provided by Osaka University through the HPCI System Research
 Project (Project ID:hp170003, hp180002).

\bibliographystyle{JHEP}
\bibliography{spectrum}

\end{document}